

\font\titolino=cmbx10
\font\tsnorm=cmr10
\font\tscors=cmti10

\font\tscorsp=cmti9
\magnification=1200

\hsize=148truemm
\hoffset=10truemm
\parskip 3truemm plus 1truemm minus 1truemm
\parindent 8truemm
\newcount\notenumber

\def\note{\advance\notenumber by 1 \footnote{$^{\the\notenumber}$}}
\def\ref#1{\medskip\everypar={\hangindent 2\parindent}#1}
\def\beginref{\begingroup
\bigskip
\leftline{\titolino References.}
\nobreak\noindent}
\def\endref{\par\endgroup}
\def\beginsection #1. #2.
{\bigskip
\leftline{\titolino #1. #2.}
\nobreak\noindent}
\def\beginappendix #1.
{\bigskip
\leftline{\titolino Appendix #1.}
\nobreak\noindent}

\nopagenumbers
\rightline{SISSA 74/94/A}
\rightline{}
\vskip 20truemm
\centerline{\titolino WORMHOLE SOLUTIONS}
\bigskip
\centerline{\titolino IN THE KANTOWSKI--SACHS SPACETIME}
\vskip 15truemm
\centerline{\tsnorm Marco Cavagli\`a}
\bigskip
\centerline{\tscorsp ISAS, International School for Advanced
Studies, Trieste, Italy}
\centerline{\tscorsp and}
\centerline{\tscorsp INFN, Sezione di Torino, Italy.}
\vfill
\centerline{\tsnorm ABSTRACT}
\begingroup\tsnorm\noindent
I present quantum wormhole solutions in the Kantowski--Sachs spacetime
generated from coupling the gravitational field to the axion and EM
fields. These solutions correspond to the classical ones found by Keay
and Laflamme and by Cavagli\`a et al.
\vfill
\hrule
\noindent
Mail Address:
\hfill\break
ISAS-International School for Advanced Studies
\hfill\break
Via Beirut 2-4, I-34013 Miramare (Trieste)
\hfill\break
Electronic mail: 38028::CAVAGLIA or CAVAGLIA@TSMI19.SISSA.IT
\endgroup
\eject
\footline{\hfill\folio\hfill}
\pageno=1
Wormholes (WHs) are classical or quantum solutions for the gravitational
field describing a bridge between two asymptotic manifolds. In the
classical case they can be interpreted as instantons describing a
tunnelling between two distant regions; on the contrary, in the quantum
case, WHs are non singular solutions of the Wheeler--de Witt (WDW)
equation with particular boundary conditions which point out that the
manifold is asymptotically euclidean [1,2]: following the Hawking--Page
prescription the wave function for large 3--metrics must reduce to the
vacuum wave function defined by a path integral over all asymptotically
euclidean metrics.

In the classical theory, pure gravity WH solutions do not exist and it
seems that only particular kinds of matter and gauge fields coupled to
gravity generate WH solutions. In particular, spherically symmetric WHs
generated from coupling the gravitational field to the axion, to the
conformal massless field and to a Yang--Mills field have been discussed
in the literature [3-8]. The magnetic monopole solution in four
dimensions with symmetry $S^2\times S^1$ has been investigated by Keay
and Laflamme [9].

On the contrary, in the quantum formalism it seems (but not proved) that
every matter field could generate WH solutions. Hawking and Page [2],
Campbell and Garay [10] and Gonz\'alez-D\'\i az [11] have discussed the
case of massless minimal coupled scalar field, Hawking and Page [2] have
also discussed the cases of massive minimal coupled and conformal scalar
field. The fact that every matter or gauge field could admit WH solutions
is very important: few years ago S. Coleman argued that the presence of a
wormhole coherent foam could drive the cosmological constant to zero
[12]; however this consideration should be very hard to support if only
particular fields generate WH solutions. Therefore it seems of crucial
importance to explore which matter or gauge fields generate WH solutions.

The aim of this letter is to present quantum WHs in the Kantowski--Sachs
(KS) spacetime, i.e. non spherically symmetric solutions of the WDW
equation. In particular I will deal with axion and EM fields.
The structure of the paper is the following: I will recall briefly
classical solutions and then I will find corresponding WDW solutions. As
we will see below, unlike the classical framework, in quantum
formalism pure gravity WH solutions do exist and I will show that the
EM field generates WH in canonical quantum gravity.

Let us consider the spacetime manifold to be described by the
KS euclidean line element:
$$ds^2=N^2(t)dt^2+a^2(t)d\chi^2+b^2(t)d\Omega_2^2\eqno(1)$$
where $\chi$ is the coordinate of the 1-sphere, $0\le\chi<2\pi$,
$d\Omega_2^2$ represents the line element of the 2-sphere and $N(t)$ is
the lapse function. We can simplify the line element choosing $N(t)=1$.
The action of our problem is
$$S_E=\int_\Omega d^4x\sqrt g\Biggl[-{M_p^2\over 16\pi}R+L(\phi)\biggr]
+\int_{\partial\Omega} d^3x\sqrt h {M_p^2\over 8\pi}{\bf K}\eqno(2)$$
where $\Omega$ is the compact four dimensional manifold described by
(1), $R$ is the curvature scalar, $L(\phi)$ is the matter lagrangian,
${\bf K}$ is the trace of the extrinsic curvature of the boundary
$\partial\Omega$ of $\Omega$ and $h$ is the determinant of the induced
metric over $\partial\Omega$.

Let us consider the case $L(\phi)=0$. Substituting (1) in (2) we
obtain ($M^2_p=1$ and we neglect some constants):
$$S_E=-\int dt\bigl[a\dot b^2+2\dot ab\dot b+a\bigr]\eqno(3)$$
where the dot represents differentiation with respect to $t$ and we have
integrated over the coordinates of the 1-sphere and 2-sphere. The problem
is then formally reduced to two degrees of freedom: $a$ and $b$. From
(3) it is easy to derive the hamiltonian constraint and the classical
equations of motion for $a$ and $b$ [13]. Thus we can easily deduce that the
only non singular and asymptotically flat solution is given by $b^2=t^2$
and $a^2=const.$, and it represents a $R^3\times S^1$ flat euclidean
space. This result is well known and not surprising: in fact one can
prove that the only solution of the Einstein equations asymptotically
flat with zero energy is the flat space [14] and that in order to allow
topology changes the eigenvalues of the Ricci tensor must be negative
somewhere on the manifold [15]. Therefore, no pure gravity WH solutions
exist in the classical euclidean frame.

If we add an axion field the action (3) becomes
$$S_E=-\int dt\bigl[a\dot b^2+2\dot ab\dot b+a+ab^2\dot\Phi^2\bigr]
\eqno(4)$$
where $\Phi$ is the axion pseudoscalar which corresponds in the euclidean
framework to an imaginary massless scalar field. From (4) we can easily
write the equations of motion and after some algebra we obtain [9]:
$$\eqalignno{&ds^2=dt^2+\bar c^2d\chi^2+(t^2+c^2)d\Omega_2^2,&(5a)\cr
&\Phi=\arctan{{t\over c}}&(5b)}$$
where $c$ is an integration constant related to the flux of the axionic
field through any $S^1\times S^2$ three--manifold and $\bar c$ is an
integration constant with dimension of length whose value will remain
arbitrary. The solution (5) was first found by Keay and Laflamme [9].
(5) is never singular and for $t\rightarrow\infty$ $b^2\rightarrow
t^2$. This asymptotic behaviour ensures that the solution can be
interpreted as a WH connecting two asymptotic flat space regions with
topology $R^3\times S^1$ and thus $(5a)$ represents a WH joining two
asymptotically $R^3\times S^1$ regions. Now let us examine the case of an
electromagnetic field. If we choose for the vector potential the ansatz
[13] (see also [9] for the dual)
$$A_\mu=A(t)\delta_{\chi\mu},\eqno(6)$$
the action (3) becomes:
$$S_E=-\int dt\biggl[a\dot b^2+2\dot ab\dot b+a-{b^2\over a}\dot A^2\biggr].
\eqno(7)$$
In this case the solution is:
$$\eqalignno{ds^2&=dt^2+\bar c^2{t^2\over
c^2+t^2}d\chi^2+(c^2+t^2)d\Omega_2^2,&(8a)\cr
A(t)&={\bar cc\over\sqrt{c^2+t^2}}&(8b)}$$
where, as in previous case, $c$ is an integration constant related to the
flux of the EM field and $\bar c$ is an integration constant with
dimension of length whose value will remain arbitrary. Again the solution
can be interpreted as a WH connecting two asymptotic flat space regions
$R^3\times S^1$. At $t=0$ the metric seems singular but this is only due
to the choice of the coordinates, that cover only half of the WH (for a
complete discussion, see [13]). Indeed, in the neighbourhood of $t=0$,
redefining the variable $\chi$, the line element becomes ($\bar c = c$)
$$ds^2=dt^2+t^2d\chi^2+c^2d\Omega_2^2.\eqno(9)$$
In the neighbourhood of $t=0$ our solution coincides with an euclidean
Kasner universe [16]. Naturally the singularity at $t=0$ can be
eliminated by going to cartesian coordinates in the $(t,\>\chi)$ plane.
This particular case of singularity removable by a different choice of
coordinates has been classified by Gibbons and Hawking [17] as a `bolt'
singularity. In the neighbourhood of $t=0$ the topology is locally
$R^2\times S^2$ with $R^2$ contracting to zero as $t\rightarrow 0$.

Now we look for the quantum solutions corresponding to the classical
solutions discussed above. To quantize the action (3) we must associate
the classical quantities $a$ and $b$ and their momenta to quantum
operators. Then it is advisable to cast the action (3) in a different
form because the lagrangian is not quadratic in the canonical momenta.
Let us introduce a change of variables in order to separate the canonical
momenta:
$$\eqalignno{a&=f(x,y),&(10a)\cr
b&=g(x,y).&(10b)\cr}$$
Substituting into the lagrangian we can require the term
proportional to $\dot x\dot y$ to vanish; if $f$ does not depend on $y$
this condition implies that the functions $f$ and $g$ fulfill
$$f{\partial g\over\partial x}+g{\partial f\over\partial x}=0.\eqno(11)$$
Analogous results can be obtained with the conditions $f(x,y)\equiv
f(y)$, or $g(x,y)\equiv g(x)$, or $g(x,y)\equiv g(y)$. From (11) we
obtain the final form for $a$ and $b$:
$$\eqalignno{a&=f\equiv f(x),&(12a)\cr
b&=g={h(y)\over f(x)}&(12b)\cr}$$
where $h(y)$ is an arbitrary function of $y$. Substituting (12) into the
lagrangian (3) we can write the classical hamiltonian in terms of the new
canonical variables $x$ and $y$ and of their canonical momenta:
$$H=-{f\over 4h^2}\Biggl[{1\over(h'/h)^2}\Pi_y^2
-{1\over(\dot f/f)^2}\Pi_x^2-4h^2\Biggr]\eqno(13)$$
where now the dot represents differentiation with respect to $x$ and the
prime denotes differentiation with respect to $y$.
Now we can quantize the system substituting quantum operators for the
classical quantities. Choosing the Laplace--Beltrami ordering for the
kinetic part, the WDW equation in the pure gravity case becomes:
$$\biggl[{\partial^2\over\partial\beta^2}-
{\partial^2\over\partial\alpha^2}-
4e^{2\beta}\biggr]\Psi(\alpha,\beta)=0\eqno(14)$$
where $f(x)=\exp[\alpha(x)]$ and $h(y)=\exp[\beta(y)]$. Now it is
straighforward to write the bounded solution of (14):
$$\Psi_\nu(f,h)=f^{\pm i\nu}K_{i\nu}(2h)\eqno(15)$$
or, in terms of the scale factors $a$ and $b$:
$$\Psi_\nu(a,b)=a^{\pm i\nu}K_{i\nu}(2ab)\eqno(16)$$
where $\nu$ is a real number and $K_{i\nu}$ is the modified Bessel function
of index ${i\nu}$ [18]. The asymptotic behaviour of the wave function
(16) for large values of the scale factor $b$ is:
$$\Psi\approx\exp(-2ab).\eqno(17)$$
This is exactly the asymptotic behaviour of a solution representing a
wormhole with spatial symmetry $S^2\times S^1$ joining two flat regions
$R^3\times S^1$ (see for instance [10]); however the behaviour of (16) is
not regular at $b\rightarrow 0$ since $\Psi$ goes like $b^{\pm i\nu}$ as
one approaches $b=0$; then (16) cannot be interpreted as a (pure gravity)
quantum wormhole but it represents an empty quantum baby universe with an
initial singularity and maximum dimension $ab\approx\nu$. We can
construct a type WH solution considering a superposition of wave
functions (16):
$$\Psi^{WH}=\int d\nu C(\nu)\Psi_\nu.\eqno(18)$$
Choosing for instance $C(\nu)=e^{i\nu\mu}$ (Fourier transform) and
substituting (16), we can easily calculate $\Psi^{WH}$ (see [10], [19]):
$$\Psi^{WH}_\mu(a,b)=e^{-2ab\cosh{[\log{a}+\mu]}}\eqno(19)$$
that it is regular and asymptotically equivalent to (3.16). Then (19)
represents a pure gravity WH.

In the same way we can discuss the case with $L(\phi)\not=0$.
If we consider the axion field the hamiltonian becomes:
$$H=-{f\over 4h^2}\Biggl[{1\over(h'/h)^2}\Pi_y^2
-{1\over(\dot f/f)^2}\Pi_x^2-4h^2+\Pi_\Phi^2\Biggr].\eqno(20)$$
In this case the hamiltonian (20) can be separated into the
gravitational and matter field parts. Then the operator ordering for the
matter do not depend on the structure of the manifold and we can quantize
the matter degree of freedom substituting:
$$\Pi_\Phi^2\rightarrow\partial_\Phi^2\eqno(21)$$
We can easily separate the equation and solve the gravitational part as
in pure gravity case. We obtain:
$$\Psi_{(\nu,\omega)}(a,b,\Phi)=
a^{\pm i\nu'} K_{i\nu}(2ab)e^{\pm i\omega\Phi}\eqno(22)$$
where $\nu'^2=\sqrt{\nu^2+\omega^2}$. Similar solutions were first found
by Campbell and Garay [10] few years ago for the massless scalar field
case; they reduce to (16) for $\omega=0$.
We note that (22) are eigenfunctions of $\Pi_\Phi$ and therefore there
is a flux of the axion field through each threedimensional surface at
$t=constant$. As in the previous case (22) cannot be interpreted as
quantum WHs but we can still construct a superposition that can
be interpreted as a WH (as in [10]).

If we consider the EM field case, analogously to the previous case the
hamiltonian can be separated into the gravitational and matter
parts. The solution corresponding to $(8)$ is:
$$\Psi_{(\nu,\omega)}(a,b,A)=K_{i\nu}(\omega a)K_{i\nu}(2ab)
e^{\pm i\omega A}.\eqno(23)$$
Again these solutions cannot be interpreted as WHs but we are able to
construct a superposition of (23) that describes a
WH. Starting from (18) we can choose for instance
$$C(\nu)=\nu\tanh(\pi\nu).\eqno(24)$$
Substituting (23) and (24) in the (18) we obtain the normalized
wave function (Kontorovich--Lebedev transform, see [19]):
$$\Psi_\omega(a,b,A)={\sqrt{16b\omega}\over\omega+2b}
e^{-a(\omega+2b)}e^{\pm i\omega A}.\eqno(25)$$
We can easily verify that (25) is regular and its asymptotic
behaviour for $b\rightarrow\infty$ coincides with (17): then
(25) can be interpreted as a WH generated by the EM field joining two
flat $R^3\times S^1$ regions.

In this paper I have examined quantum WH solutions in the KS spacetime.
In particular I have dealt with axion and EM fields. In pure gravity
case it is well known that WH type solutions do not exist in the
classical euclidean frame; on the contrary in the quantum frame, as we
have seen, solutions of the WDW equation that can be interpreted as WH
wave functions do exist. In the axion case the quantum solutions are
similar to the solutions for the massless scalar field discussed by
Campbell and Garay. In the EM field case the quantum WH solutions are the
first example of quantum WH generated by the coupling between the
canonical quantized gravitational field and the EM field. I have then
proved that the EM field generates WH in canonical quantum
gravity.

I am very grateful to Prof. V. de Alfaro and L.J. Garay for interesting
discussions. I also thank Prof. J. Nelson.
\vfill\eject
\beginref
\ref [1] L.J. Garay, {\tscors Phys. Rev.} D {\bf 48}, 1710 (1993).
\ref [2] S.W. Hawking and D.N. Page, {\tscors Phys. Rev.} D {\bf
42}, 2655 (1990).
\ref [3] S.B. Giddings and A. Strominger, {\tscors
Nucl. Phys.} B {\bf 306}, 890 (1988).
\ref [4] R.C. Myers, {\tscors Phys. Rev.} D {\bf 38}, 1327 (1988).
\ref [5] J.J. Halliwell and R. Laflamme, {\tscors Class.
Quantum Grav.} {\bf 6}, 1839 (1989).
\ref [6] D.H. Coule and K. Maeda, {\tscors Class. Quantum
Grav.} {\bf 7}, 955 (1990).
\ref [7] S.W. Hawking, {\tscors Phys.  Lett.} B {\bf 195}, 337 (1987).
\ref [8] A. Hosoya and W. Ogura, {\tscors Phys. Lett.} B {\bf
225}, 117 (1989).
\ref [9] B.J. Keay and R. Laflamme, {\tscors Phys. Rev.} D
{\bf 40}, 2118 (1989).
\ref [10] L.M. Campbell and L.J. Garay, {\tscors
Phys. Lett.} B {\bf 254}, 49 (1991).
\ref [11] P.F. Gonz\'alez--D\'\i az, P.F., {\tscors Preprint}
IMAFF-RC-02-93, {\bf gr-qc}/9306031 (1993).
\ref [12] S. Coleman, {\tscors Nucl. Phys.} B {\bf 310}, 643 (1988).
\ref [13] M. Cavagli\`a, V. de Alfaro and F. de Felice, {\tscors
Phys. Rev.} D {\bf 49}, 6493 (1994).
\ref [14] R. Schoen and Shing-Tung Yau, {\tscors Commun. Math.
Phys.} {\bf 79}, 231 (1982).
\ref [15] J. Cheeger and D. Gromoll, {\tscors Ann. Math.} {\bf
96}, 413 (1972).
\ref [16] C.W. Misner, K.S. Thorne and J.A. Wheeler, {\tscors
Gravitation}, W. H. Freeman and Company, New York 1973.
\ref [17] G.W. Gibbons and S.W. Hawking, {\tscors Commun. Math.
Phys.} {\bf 66}, 291 (1979).
\ref [18] I.S. Gradshteyn and I.M. Ryzhik, {\tscors Table of Integrals,
Series and Products}, Academic Press, San Diego 1980.
\ref [19] Bateman Manuscript Project, {\tscors Tables of Integral
Transforms}, 1954, Mc. Graw--Hill Book Company, New York.
\endref
\vfill\eject
\bye